\documentclass[aps,prd,nopacs,nofootinbib,notitlepage,superscriptaddress]{revtex4-1}

\usepackage{amsfonts}
\usepackage{amsmath}
\usepackage{xcolor}
\usepackage{bm}
\usepackage{bbm}
\usepackage{graphicx}
\usepackage{fancybox}
\usepackage{comment}
\usepackage{multirow}
\usepackage{tipa}
\usepackage{longtable}
\definecolor{refs}{RGB}{245,156,74}
\usepackage[colorlinks=true,hyperfootnotes=true,citecolor=cyan]{hyperref}

\usepackage{enumitem}

\newcommand{\be}{\begin{equation}}
\newcommand{\ee}{\end{equation}}
\newcommand{\ba}{\begin{eqnarray}}
\newcommand{\ea}{\end{eqnarray}}
\newcommand{\bs}{\begin{subequations}}
\newcommand{\es}{\end{subequations}}

\newcommand{\e}{\mathrm{e}}
\newcommand{\ie}{\text{\textschwa}}

\newcommand{\diff}{\textrm{d}}

\newcommand{\lp}{\left(}
\newcommand{\rp}{\right)}

\newcommand{\tomi}[1]{{\textcolor{red}{ #1}}}

\begin{document}

\title{Euclidean teleparallel relativity and black hole partition functions}

\author{Jose Beltr{\'a}n Jim{\'e}nez}
\email{jose.beltran@usal.es}
\affiliation{Departamento de F\'isica Fundamental and IUFFyM, Universidad de Salamanca, E-37008 Salamanca, Spain}

\author{Tomi S. Koivisto}
\email{tomi.koivisto@ut.ee}
\address{Laboratory of Theoretical Physics, Institute of Physics, University of Tartu, W. Ostwaldi 1, 50411 Tartu, Estonia}
\address{National Institute of Chemical Physics and Biophysics, R\"avala pst. 10, 10143 Tallinn, Estonia}

\begin{abstract}

The Euclidean path integral approach to quantum gravity is conventionally
formulated in terms of the Einstein-Hilbert-York-Gibbons-Hawking action, 
which requires suitable subtractions to produce the correct black hole partition function. However, there is a 
unique, canonical teleparallel reformulation which reproduces the same results without subtractions or other 
ambiguities. This is verified in the case of a black hole with or without an electric or a    
magnetic charge and in a background with or without a cosmological constant. 
Moreover, a new quasilocal prescription is proposed and tested, where the black hole partition function is determined solely by the horizon boundary term, yielding the correct Helmholtz free energy without counterterms.

%The results may be further improved in the quasilocal prescription which 
%reduces the canonical teleparallel action to a surface integral over the horizon.

\end{abstract}

\maketitle

\section{Introduction}
\label{sec1}

In the well-known derivation of the black entropy from the Euclidean action gravity by Gibbons and Hawking \cite{Gibbons:1976ue}, the boundary is decisive. 
The Einstein-Hilbert action is supplemented with the York-Gibbons-Hawking term \cite{York:1972sj} in the standard formulation of General Relativity (referred to as 
GR from now on), and in the case of the Schwarzschild black hole, it is the latter, together with an extra subtraction term, that determines the nonzero %- and remarkably, \jose{{\it the correct}}- 
result  for the black hole entropy in the Euclidean approach \cite{Gibbons:1976ue}. More general cases often require the suitable adjustments of the prescription for the boundaries and for the subtractions \cite{Gibbons:1994cg}.  

Teleparallel formulations of gravity differ from GR in that they do not require an additional boundary term for the consistency of the variational principle, and the potential advantage of a teleparallel formulation in Euclidean quantum gravity and in related frameworks has been pointed out in some recent works, 
e.g \cite{BeltranJimenez:2017tkd,Oshita:2017nhn,BeltranJimenez:2018vdo,Hammad:2019oyb,Heisenberg:2022nvs,Koivisto:2022oyt,Erdmenger:2022nhz,Erdmenger:2023hne,Fiorini:2023axr,Krssak:2023nrw,Kuntz:2024qgs,Krssak:2024vzo}. 
However, the consistency of the variational principle is achieved at the expense of realising local symmetries only up to total derivatives. The boundary term in teleparallel gravity actions depends upon a frame field which is not determined by field equations. There is always an {\it infinite} number of frames that would produce {\it any arbitrary} result from a computation in the Euclidean approach\footnote{For example, the recent paper  \cite{Kuntz:2024qgs} adopts the prescription \cite{Lucas:2009nq,Krssak:2015rqa, Emtsova:2021ehh} which only partially fixes the frame, and only at infinity. In general, this leaves 16 free functions (6 if restricting to metric frames, 4 if restricting to holonomic frames), only subject to one scalar constraint. The results depend on those chosen functions and are thus completely arbitrary. For some fortunate choices of the free functions the authors obtain the desired results, for some others, they have to resort to ad hoc counterterms to adjust the results \cite{Kuntz:2024qgs}.}. 
Rather than offering a robust alternative to GR, teleparallel implementations such as in Refs.\cite{BeltranJimenez:2017tkd,Oshita:2017nhn,BeltranJimenez:2018vdo,Hammad:2019oyb,Heisenberg:2022nvs,Koivisto:2022oyt, Erdmenger:2022nhz,Erdmenger:2023hne,Fiorini:2023axr,Krssak:2023nrw,Kuntz:2024qgs,Krssak:2024vzo} 
appear to leave the theory mired in arbitrariness stemming from uncontrolled reference frame freedom.
%appear to represent a step backward,  regressing into arbitrariness stemming from uncontrolled reference frame freedom.
%A way past this impasse 

The way out of this morass has been recently proposed in the extension of GR dubbed G$_\parallel$R \cite{Gomes:2023hyk}, in the framework of general teleparallel gravity \cite{BeltranJimenez:2019odq}. In contrast to popular models of teleparallel gravity, G$_\parallel$R is a predictive theory. We will show that it not only resolves the core ambiguity that has long plagued teleparallel gravity, but may even offer a concrete improvement over standard GR.
%While the Euclidean derivation of the Schwarzschild black hole entropy was already sketched in \cite{BeltranJimenez:2019bnx}, and some of the thermodynamical aspects were more thoroughly discussed in \cite{Gomes:2022vrc}, these treatments did not fully engage with the Euclidean formalism. 
In this Letter, we elaborate on the partition function approach in G$_\parallel$R and apply it to charged black holes and $\Lambda$-backgrounds. 
That these computations reproduce the correct results without invoking any counterterms is a strong and nontrivial confirmation of the theory’s viability, lending weight to the conjecture that quantum G$_\parallel$R may be renormalisable, or even finite.

This Letter is organised as follows. We recall the G$_\parallel$R action in Sec.\ref{sec2gr} and the relevant solution in \tomi{Sec.}\ref{sec3bh}. The black hole partition in the saddle point approximation to the Euclidean approach is reviewed in Sec.\ref{sec4eqg}, and its (quasi)localised application is proposed. We then check the results in de Sitter space (Sec.\ref{sec5ds}), with a charged black hole (Sec.\ref{sec6rn}), with a black hole in anti-de Sitter background (Sec.\ref{sec7sads}), before making the conclusion (Sec.\ref{sec8conc}).
%we arrive at the conventional results but now without the conventional regularisations implemented with additional \tomi{counterterm}s. 
%that the conjecture perhaps provides a potentially improved description from the thermodynamical perspective. 

\section{G$_\parallel$R}
\label{sec2gr}

A point of departure is the realisation that even in a non-trivial metric background $g_{\mu\nu}$, there should exist a frame wherein the induced phase space structure obeys the canonical commutation relations \cite{Koivisto:2018aip} and the relational observables may thus be defined unambiguously \cite{Gomes:2022vrc}.
%A point of the departure is the realisation that even in the presence of a non-trivial metric field $g_{\mu\nu}$, there should exist some frame of reference whose induced coordinates and momenta satisfy the canonical commutation relations \cite{Koivisto:2018aip}. 
In general, this may not be achieved with a coordinate frame, but it requires 4 vectors  $\ie_a^\mu$ to span a General Linear (GL) frame. We denote the inverse frame $\e^a_\mu$, so that $\e^a_\mu\ie_b^\mu=\delta^a_b$ and $\e^a_\mu\ie_a^\nu=\delta^\nu_\mu$. The diffeomorphism (Diff) {\it gauge symmetry} is thus extended by the GL {\it frame symmetry} \cite{Koivisto:2022oyt,Gomes:2023hyk}. Whereas the gauge symmetry is a redundancy of description, the frame transformations ($\ie_a^\mu \rightarrow A_a{}^b\ie_b^\mu$, where $A_a{}^b \in$ GL) modify the symplectic structure associated with the equivalent dynamics and, in particular, change the boundaries of the action. In G$_\parallel$R, the $\ie_a^\mu$ is deduced from the metric and the matter content (the latter being specified by the matter energy tensor $T_{\mu\nu} = (2/\sqrt{-g})\delta (\sqrt{-g}L_M)/\delta g^{\mu\nu}$ where $L_M$ is a Lagrangian for matter fields), rather than introduced as an extra assumption.  

For the present practical purpose, the $\ie_a^\mu$ can be regarded as a calculational tool which determines the action of gravity in what we will define as the canonical frame \cite{BeltranJimenez:2019bnx}. The quartet $\ie_a^\mu$ induces the frame-compatible covariant derivative $\nabla_\mu$ with the coefficients $\Gamma^\alpha{}_{\mu\nu} = \ie_a^\alpha{}\e^a_{\nu,\mu}$. {The relevant quantity is the tensorial difference of the frame-compatible $\Gamma^\alpha{}_{\mu\nu}$ and the metric-compatible Levi-Civita $\mathring{\Gamma}^\alpha{}_{\mu\nu}$ we denote as\footnote{
The $X^\alpha{}_{\mu\nu}$ is sometimes called the ``distortion" tensor. Since the connection $\Gamma^\alpha{}_{\mu\nu}$ is by construction flat, the framework admits teleparallel geometrical interpretations \cite{BeltranJimenez:2019esp}, which have been extensively discussed in the literature. 
In this Letter, however, we instead focus on {\it relativity}, a comparatively unexplored aspect of teleparallel relativity \cite{Koivisto:2022nar}.},
\be
X^\alpha{}_{\mu\nu} \equiv \Gamma^\alpha{}_{\mu\nu} - \mathring{\Gamma}^\alpha{}_{\mu\nu}\,, \quad  V^\mu \equiv 2X^{[\mu}{}_\alpha{}^{\alpha]}\,,   
\quad X \equiv - 2X^\alpha{}_{[\alpha\lvert\beta\rvert}X^\beta{}_{\gamma]}{}^\gamma\,. 
\ee
%thus admittingHowever, here it is only relevant to define the vector $V^\mu$, 
%\be
%V^\mu = 2Q^{[\mu\alpha]}{}_\alpha 
%+ 2T^\alpha{}_{\mu\alpha}\,, \quad \text{where} \quad T^\alpha{}_{\mu\nu}  =  2\Gamma^\alpha{}_{[\mu\nu]}\,, \quad 
%Q_\alpha{}^{\mu\nu}  =  -\nabla_\alpha g^{\mu\nu}\,, 
%\ee
%\cite{Gomes:2023hyk}.
%In terms of the metric $g_{\mu\nu}$ and a frame-compatible connection $\Gamma^\alpha{}_{\mu\nu} = \ie_a^\alpha{}\e^a_{\nu,\mu}$, we define the two tensors as follows
%torsion and non-metricity tensors as follows
%and it is convenient tointroduce the notations for the traces of these tensors,
%\be
%T_\mu = T^\alpha{}_{\mu\alpha}\,, \quad 
%Q_\alpha = g_{\mu\nu}Q_\alpha{}^{\mu\nu}\,, \quad
%q^\mu = Q_\alpha{}^{\mu\alpha}\,, \quad 
%V^\mu =  Q^\mu - q^\nu + 2T^\mu\,.
%\ee
The quadratic Diff-scalar $X$  %can be defined in terms of the ,
%\be
%\quad \text{where} \quad  X^\alpha{}_{\mu\nu} = \frac{1}{2}Q^\alpha{}_{\mu\nu} - Q_{(\mu\nu)}{}^\alpha  + \frac{1}{2}T^\alpha{}_{\mu\nu} -  T_{(\mu\nu)}{}^\alpha\,,
%\ee
is} selected by the symmetry requirement that $X$ changes by a total derivative under an arbitrary local GL transformation  \cite{BeltranJimenez:2019odq}. From these definitions it follows that $X = \mathring{R} + \mathring{\nabla}_\mu V^\mu$, where the ring denotes metric variables, $\mathring{R}$ thus being the Ricci scalar computed from $g_{\mu\nu}$. The action 
\be
I_X[g,\e]=\frac{m_P^2}{2}\int \diff^4 x \sqrt{-g}X,
\ee
where $m_P$ is the Planck mass, reduces to Einstein's GR \cite{https://doi.org/10.1002/andp.19163540702} in the {\it coincident gauge} $\nabla_\mu \overset{*}{=} \partial_\mu$, since then $X^\alpha{}_{\mu\nu} \overset{*}{=} -\mathring{\Gamma}^\alpha{}_{\mu\nu}$. %\cite{BeltranJimenez:2022azb}\josec{Is this the relevant reference here?}. %For this reason we called it the relativistic completion, 
Due to the enhanced symmetry GL$\times$Diff, the equations can be considered in an arbitrary frame whilst they remain generally covariant. In a {\it canonical frame}, wherein the energy of the frame field vanishes,  the field equations can be written as \cite{Gomes:2023hyk}
\be \label{fieldeq}
\mathring{\nabla}_\alpha H_a^{\alpha\mu} = \sqrt{-g}T^\mu{}_\nu\ie_a^\nu\,, \quad \text{where} \quad
%H_a^{\mu\nu} = -\sqrt{-g}m_P^2\ie_a^\alpha\lp Q^{[\mu\nu]}{}_\alpha \tomi{ + } \delta^{[\mu}_\nu V^{\nu]} + \frac{1}{2}T_\alpha{}^{\mu\nu} - T^{[\mu\nu]}{}_\alpha \rp\,. 
{H_a^{\mu\nu} = -\sqrt{-g}m_P^2\ie_a^\alpha\lp X^{[\mu}{}_\alpha{}^{\nu]} + 2\delta^{[\mu}_\nu V^{\nu]}\rp}\,. 
\ee 
{This equation can be taken as the definition of a canonical frame.} 

If we define an {\it inertial frame} as the equivalence class of solutions for which the local energy-momentum of the metric vanishes, it follows that $X=0$ in an inertial frame. To see that, it is sufficient to notice that the energy-momentum of the metric is given by an expression $G^\mu{}_\nu$ whose trace is $G^\mu{}_\mu = 3X$ \cite{Gomes:2023hyk}. In the coincident gauge $\overset{*}{G}{}^\mu{}_\nu$ reduces to the Einstein pseudotensor, whereas in an inertial frame we have by definition $G^\mu{}_\nu=0$ and thus $X=0$ in an inertial frame. There exists also what could be called the zero-energy frame or the {\it Levi-Civita frame} $\mathring{\nabla}_\alpha H_a^{\alpha\mu} \rightarrow 0$, wherein $G^\mu{}_\nu$ reduces to the pseudo-Riemannian Einstein tensor $G^\mu{}_\nu \rightarrow -m_P^2\mathring{G}^\mu{}_\nu$. It is worth noticing that an inertial frame is also canonical if $\Gamma^\alpha{}_{[\mu\nu]}=0$. The Noether-Wald charges in teleparallel gravity were derived in \cite{BeltranJimenez:2021kpj}, and in particular, the energy $E$ is given by the integral $E=\oint H_0^{\mu\nu}\diff x_\mu\wedge\diff x_\nu$ over a closed surface. %\footnote{For a dogmatic denial of this result, see \cite{newgolovnev}. (The author also invents ``a very erroneous claim'' which he supposes as the starting point for the deniable Geometrical Trinity of Gravity; see however the original paper \cite{BeltranJimenez:2019esp}.)}.
The flux $H_0^{\mu\nu}$ is determined uniquely, since 1) all the above definitions are Diff-covariant, 2) the frame is fixed by (\ref{fieldeq}) and 3) a frame contains only one vector $\ie_{a=0}^\mu$ which is time-like, $g_{\mu\nu}\ie_0^\mu\ie_0^\nu <0$. 

\section{Classical black hole solution}
\label{sec3bh}

In the following, we shall restrict to spherically symmetric metrics of the form 
\be \label{metric}
\diff s^2 = -\big(1 - f(r)\big)\diff t^2 + \frac{\diff r^2}{1 - f(r)} + r^2\diff\Omega^2,
\ee
where $\diff\Omega^2 = \diff\theta^2 + \sin^2\theta\diff\phi^2$ is the angular part. Let us define the short-hand  $h = (1-f)^{-1}f$. An inertial frame for the metrics (\ref{metric}) is spanned by the 4 vectors
\bs
\label{frame}
\ba
%\ie_0  & = &  \lp t - \frac{1}{2}\int h(r)\diff r\rp\partial_t \quad \text{where} \quad h = (1-f)^{-1}f\,, \label{timecomponent} \\
\ie_0  & = & \partial_t\,, \label{timecomponent} \\
%\ie_1  & = &  r\sin^2\theta\cos^2\phi\lp h\partial_t + \partial_r\rp + \sin\theta\cos\theta\cos^2\phi\partial_\theta
%- \sin\phi\cos\phi\partial_\phi\,, \\
\ie_1  & = &  \sin\theta\cos\phi\lp - h\partial_t + \partial_r\rp + r^{-1}\cos\theta\cos\phi\partial_\theta - (r\sin\theta)^{-1} \sin\phi\partial_\phi\,, \\
%\ie_2  & = &  r\sin^2\theta\sin^2\phi\lp h\partial_t + \partial_r\rp + \sin\theta\cos\theta\sin^2{\phi}\partial_\theta
%+ \sin\phi\cos\phi\partial_\phi\,, \\
\ie_2  & = &  \sin\theta\sin\phi\lp - h\partial_t + \partial_r\rp + r^{-1}\cos\theta\sin\phi\partial_\theta + (r\sin\theta)^{-1} \cos\phi\partial_\phi\,, \\
%\ie_3  & = &  r\cos^2\theta\lp h\partial_t + \partial_r\rp - \sin\theta\cos\theta\partial_\theta\,, 
\ie_3  & = &  \cos\theta\lp - h\partial_t + \partial_r\rp - r^{-1}\sin\theta\partial_\theta\,.
\ea
\es
as can be checked by direct substitution. Thus, (\ref{metric},\ref{frame}) is a representative of an equivalence class of inertial frames related by a residual global GL symmetry, as well as the usual local Diff symmetry. It follows that $T^\alpha{}_{\mu\nu}=2\ie^\alpha_a\e^a_{[\nu,\mu]}=0$,
and thus this frame is also canonical\footnote{In geometrical terms, the corresponding affine geometry is ``torsion-free'' and thus ``symmetrically teleparallel''  \cite{BeltranJimenez:2019esp}. In cosmological spacetimes, the canonical and inertial frames differ \cite{Gomes:2023hyk}, which suggests that the mismatch may reflect a dynamical aspect of the metric.}. By straightforward computations, we obtain that now 
\be
V = {-\frac{2f/r+f'}{1-f}\partial_t - (2f/r+f')\partial_r}\,,
\ee
is a null vector, $V^\mu V_\mu=0$, and that the energy enclosed within the radial distance $r$ is 
\be \label{energy}
E(r)=4\pi m_P^2 rf(r).
\ee
In \cite{Gomes:2022vrc} the equivalent results were derived in the coincident gauge $\ie_a \overset{*}{=}\delta_a^\mu\partial_\mu$ using a different coordinate system. 

When Wick rotating to imaginary time $\tau = -it$, the metric acquires the Euclidean signature, 
$\diff s^2 = (1 - f(r))\diff \tau^2 + (1 - f(r))^{-1}\diff r^2 + r^2\diff\Omega^2$, and the 
frame becomes complex, since $\partial_t =  -i\partial_\tau$. In the following, we will work in these Euclidean coordinates. {Let us notice that, despite having a complex frame, we will not generate an imaginary part for the Euclidean action.} 

\section{Euclidean quantum gravity and the black hole entropy}
\label{sec4eqg}

The proposal is that the canonical partition function in terms of the Euclidean action $I_E$ which could involve some matter fields $\psi$,  
\be \label{partition}
Z = \int\mathcal{D} g \int \mathcal{D}{\psi} e^{-I_E[g,\e,\psi]} \approx e^{-I_E[g_{\text{classical}},\e,\psi_{\text{classical}}]}\,,
\ee
is evaluated in an inertial frame. The quantum theory thus differs from GR already at the leading saddle point approximation wherein only the classical solution contributes to the path integral. An immediate consequence of the proposal is that the gravitational bulk action $I_X$ for arbitrary quantum excitations identically vanishes. This is in line with the view, commonly accepted in e.g. the context of holography, that there are no local degrees of freedom in quantum gravity, and with the gauge theory of gravity wherein spacetime emerges via a symmetry breaking \cite{Koivisto:2025ryb}. We have nevertheless obtained nontrivial physics from the classical gravity theory. 
This is because the physics emerges only in relation to an observer, whereby independent variations of the metric field and the frame field yield nontrivial field equations for the former.
The presence of an observer implies a classical background, within which variations of the metric can be treated as occurring relative to that background frame, thereby giving rise to dynamics.
%This is because the physics is interpreted from the perspective of an observer, in which case independent variations of the metric field and the frame field result in nontrivial field equations for the former.  
%\tomi{The presence of an observer implies a classical background, within which variations of the metric can be treated as occurring relative to that background frame, thereby giving rise to dynamics.}
%An observer implies a classical background, and the variations of the metric can then be considered to take place in the classical background frame and therefore lead to dynamics. 
From this perspective, when expanding the partition function (\ref{partition}), we are lead to the usual perturbative, semi-classical approach, with the Euclidean GR action now replaced by 
\be \label{conjecture}
I_E = -\int\diff^3 x\sqrt{g} V^\mu n_\mu - \int\diff^4 x \sqrt{g}\lp \mathring{R} + L_M \rp\,. %\overset{\text{conjecture}}{\rightarrow}  \int\diff^3 x\sqrt{g} V^\mu n_\mu\,. 
\ee 
In thermodynamical equilibrium at the temperature $\beta^{-1}$, the energy is given by the canonical partition function $Z$ as $E=-(\log Z)_{,\beta}$ and the entropy is $S=\log Z - \beta(\log Z)_{,\beta}$. In the leading approximation (\ref{partition}), the Helmholtz free energy $F$ is directly related to the Euclidean action, $\beta F = -\log{Z} \approx I_E$. As G$_\parallel$R provides a local expression for the energy $E$, it seems natural to speculate that the action could also describe the
free energy $F$ of a localised subsystem. In concrete terms, the partition function of a black hole would then be given solely by the first term in (\ref{conjecture}), evaluated at the event horizon. %We then interpret a black hole as a compact object enclosed by the horizon 
By including the bulk integral over the whole Euclidean section, we would instead be studying a quite different system: a (typically infinite) bulk spacetime which includes the black hole. {This is the conventional approach, wherein one only takes into account the boundary at infinity. Note, however, that for arbitrary off-shell fluctuations, the horizon is also a boundary due to the conical singularity that is present even for infinitesimal deviations from thermal equilibrium}. In the following, we shall check the predictions of G$_\parallel$R in 1) the conventional approach considering the total spacetime and in 2) the (quasi)localised approach to the horizon thermodynamics.

First, we check the simple case of the Schwarzschild black hole with mass $m_S$. Then $f=r_S/r$, where $r_S=m_S/4\pi m_P^2$. Let's evaluate 
\be
I_E [g_{\text{classical}},\e_{\text{inertial]}} 
= -\frac{m_P^2}{2}\int\diff^3 x\sqrt{g} V^\mu n_\mu =
 \frac{m_P^2}{2}\beta r^2\int\diff^2\Omega \frac{r_S}{r^2} = 2\pi m_P^2 \beta r_S = \frac{1}{2}(m_P\beta)^2\,.  \label{Sresult}
\ee
We prescribed a normal vector $n$ to a constant-$r$ hypersurface as {$n = (1-f)^{\frac{1}{2}}\partial_r$}, the period of $\tau$ as $\beta = m_S/m_P^2 = 4\pi r_S$, and chose the hypersurface at $r=r_S$. Now $E=m_S$, $S=I_E=(8\pi m_P^2)A/4$, where $A=4\pi r_S^2$. In this case, there is no difference between 1) and 2), since the bulk integral is vanishing {and the
boundary term is independent of $r$}.  

\section{de Sitter space}
\label{sec5ds}

The static patch of de Sitter space is covered by the coordinates (\ref{metric}) with $f=(r/r_{\Lambda})^2$, where $r_{\Lambda}=\sqrt{3/\Lambda}$. Simple geometrical
arguments give $\beta=2\pi r_{\Lambda}$, $A = 4\pi r^2_{\Lambda}$, $V=4 \pi r^3_{\Lambda}/3$, so that the total volume of the 4-dimensional Euclidean patch is $N=\int \diff^4 x\sqrt{g} = \beta V = 8\pi^2 r^4_{\Lambda}/3$. Since the Euclidean de Sitter space has no boundary, the full action reduces to the
simple bulk integral,
\be \label{GRbulk}
I_E= -\frac{m_P^2}{2}\int \diff^4 x\sqrt{g}\lp \mathring{R}-2\Lambda\rp = -\frac{m_P^2}{2}\int \diff^4 x\sqrt{g}\lp \frac{12}{r^2_{\Lambda}}-\frac{6}{r^2_{\Lambda}}\rp = - (8\pi m_P^2)\frac{A}{4}\,.
\ee
This is the same result as has been established in GR.

In the thermodynamical interpretation, we should take into account that the cosmological constant is associated with the pressure $p=-m_P^2\Lambda$. Thus the equation of state is $S=-24\pi^2m_P^4/p$, and a natural thermodynamic potential for the system would be the Gibbs free energy $G(\beta,p)$. We consistently obtain that $\partial G/\partial \beta = \beta^{-2}S$, $\partial G/\partial p = V$, and see that the enthalpy $E+pV$ vanishes and thus the energy is $E= m_P^2\Lambda V$, in agreement with the energy charge (\ref{energy}) in G$_\parallel$R. The Gibbs free energy $G = \mu N$ implies the presence of a chemical potential $\mu = -m_P^2\Lambda/2\pi$ of de Sitter space, when $N$ is, following \cite{Gibbons:1979xm}, considered as the 4-volume $N=\beta V$. 

Thus, the standard approach leads to the identification of the $I_E = \beta G$ in terms of the Gibbs free energy. According the quasilocal approach 2), we can instead identify $Z$ as 
the canonical partition which yields the Helmholtz free energy $F(\beta,V)$ once we ignore the bulk (\ref{GRbulk}). Since there is no boundary in this case, the free energy vanishes, as it indeed does for the de Sitter space, $F=E-TS=0$.  

\section{Charged black hole}
\label{sec6rn}

For a black hole with the charge $q$, we have $f=r_S/r - (q/r)^2$, and there are two horizons at $r_\pm = r_S/2 \pm \sqrt{r_S^2/4 - q^2}$. We are interested in the outer, event horizon whose temperature is $\beta^{-1} = (1/r_+ - q^2/r_+^3)/4\pi$.  
%The temperatures of the horizons are $\beta_{\pm}^{-1} = \lp r_S/r_{\pm}^2-2q^2/r_{\pm}^3\rp/4\pi$.
Again $\mathring{\nabla}_\mu V^\mu = \mathring{R} = 0$,  
reflecting the scale-invariance of the electromagnetic source. In fact, $V^\mu$ is independent of the charge $q$. Thus, according to the quasilocal description 2), the gravity contribution to the partition function is given by the boundary term
\be \label{RNboundary}
-\frac{m_P^2}{2}\int\diff^3 x\sqrt{g} V^\mu n_\mu = \frac{1}{2}\beta m_S\,. 
\ee
To contrast with this simple and direct derivation of the partition function $Z \approx e^{-\beta m_S/2}$, we also re-derive the standard result permitting us to highlight some unsatisfactory elements in the derivation. 
%\jose{This computation of the gravity contribution to the partition function is so simple and direct that it may even raise some suspicions. In order to better appreciate the beauty of the above computation, we will now re-derive the classical result for comparison purposes. Moreover, going through the standard derivation will also permit us to highlight certain unsatisfactory elements that remain in that derivation and which are resolved within our simpler approach.} 

In the standard approach, one also takes into account the Maxwell field $F_{\mu\nu} = 2A_{[\nu,\mu]}$. 
For generality, consider also a possible magnetic field and a possible axionic extension of the electrodynamic theory. The Maxwell field and its dual $\tilde{F}_{\mu\nu} = \frac{1}{2}\epsilon_{\mu\nu\rho\sigma} F^{\rho\sigma}$ are then described by the non-vanishing components 
\bs
\label{EMfield}
\ba 
%F_{tr} & = & -F_{rt} = \frac{q_e}{r^2}\,, \quad F_{\theta\phi} = -F_{\phi\theta} = -q_b\sin\theta\,, \\
%\tilde{F}_{tr} & = & -\tilde{F}_{tr}  = \,. 
F_{\mu\nu}\diff x^\mu\wedge\diff x^\nu & = & \frac{q_e}{r^2}\diff \tau\wedge \diff r - q_b\sin\theta\diff \theta\wedge\diff \phi\,, \\
\tilde{F}_{\mu\nu}\diff x^\mu\wedge\diff x^\nu & = & -\frac{q_b}{r^2}\diff \tau\wedge\diff r + q_e\sin\theta\diff \theta\wedge\diff \phi\,, \
\ea
\es
such that $8\pi m_P^2q^2 = q_e^2 + q_b^2$. The standard action for this field including an axion $\alpha$ would be 
\be \label{maxwell1}
I_{M} = \frac{1}{16\pi}\int \diff^4 x\sqrt{g}\lp F_{\mu\nu} F^{\mu\nu} + \alpha\tilde{F}_{\mu\nu} F^{\mu\nu} \rp=  \frac{1}{8\pi}\int\diff \tau \diff r \diff^2\Omega r^2
\lp \frac{q^2_b}{r^4} - \frac{q^2_e}{r^4}\rp = \frac{1}{2}\int\diff \tau\diff r \frac{q^2_b-q^2_e}{r^2} \,, 
\ee
and thus, we obtain for the field configuration (\ref{EMfield}) that
\be \label{RN}
I_E = \frac{1}{2}\beta\lp m_S - \frac{q^2_e}{r_+} + \frac{q^2_b}{r_+}\rp\,. 
\ee 
Switching off either $q_e$ or $q_b$, we recover the cases studied separately in \cite{Hawking:1995ap}. There the difference of the signs was explained to be due to the nature of boundary conditions on the electric and the magnetic sectors. At infinity, one cannot fix the electric charge but must consider its potential, whereas one can impose a magnetic charge at infinity. Therefore, it would be appropriate to consider the canonical partition function $Z(\beta,q_b)$ in the case of a magnetic field, but in the case of an electric field, instead the grand canonical partition function $Z(\beta,U_e)$, where $U_e$ is the electric potential. In any case, we could write the first law for the coupled system in the form
\be
\diff m_S = \beta^{-1}\diff S + U_e\diff q_e + U_b\diff q_b\,,  
\ee
and deduce from the expression for the entropy,
\be
S = (8\pi m_P^2)A/4 = 2\pi^2 m_P^2\lp r_S + \sqrt{r_S^2 - 4q^2}\rp^2\,, \label{RNentropy}
\ee
that $U_e = -q_e/r_+$ and $U_b = -q_b/r_+$. As suggested in \cite{Hawking:1995ap}, it is possible to consider the result (\ref{RN}) in terms of a charge-projected partition function, or alternatively, adjust the result by adding a boundary term to the electromagnetic action $I'_M$.  
%For curiosity, we consider an alternative prescription wherein the axion turns out to be crucial. 
%The action (\ref{maxwell1}) is divergent in the interior. The electromagnetic energy $E = -q^2/2r$ enclosed within the radius $r$ is arbitrarily large as one considers an infinitesimal radius $r$, but the $E$ is finite for any finite $r$ \cite{Gomes:2022vrc} . Thus we expect that there exists a regular version of the action. 
%To find it, we first note that the action (\ref{maxwell1}) can be rewritten as
Let us adopt the latter approach, and write\footnote{We note that, with the boundary term included as in (\ref{maxwell1b}), the canonical Noether charge associated with the electric field would be $-q_e$ rather than $q_e$}. 
\be \label{maxwell1b}
I_M \rightarrow I_M -  \frac{1}{4\pi}\int\diff^4 x \sqrt{g}\mathring{\nabla}_\mu( F^{\mu\nu}A_\nu + \alpha\tilde{F}^{\mu\nu}A_\nu )\,.    
\ee
Now the action appears to be gauge-dependent. However, at least in the case at hand, regularity dictates the suitable form of the gauge potential that reproduces the field (\ref{EMfield}), 
\be \label{EMsolution}
A_\mu\diff x^\mu = i \lp \frac{q_e}{r} +  U_e\rp\diff \tau + q_b\lp \cos\theta - 1\rp\diff\phi\,,  
%A^\pm_\mu = q_e \lp \frac{1}{r^2} - \frac{1}{r_\pm^2}\rp t_{,\mu} + q_b\cos\theta\phi_{,\mu}\,. 
\ee
where $U_e$ is the electric potential at infinity \cite{Hawking:1995ap} with the value $U_e=-q_e/r_+$ we deduced above. Evaluating (\ref{maxwell1b}) for this solution we obtain $I_M \rightarrow I_M + \beta\lp q_e - \alpha q_b\rp\frac{q_e}{r_+}$, and thus, by
setting $\alpha=0$, we arrive at the adjusted action
\be \label{RN2}
I_E \rightarrow \frac{1}{2}\beta\lp m_S + \frac{q^2_e}{r_+} + \frac{q^2_b}{r_+}\rp\,. 
\ee
When taking into account that the energy (\ref{energy}) at infinity is $E(r=\infty) = m_S$, and assuming that only the black hole contributes to the entropy $\beta^{-1}S = (m_S + U_e q_e + U_b q_b)/2$, {the result (\ref{RN2}) can be interpreted} as the canonical partition function of the \tomi{entire} spacetime, {with} $I_E = \beta F$. {This is the standard interpretation following Hawking and Ross \cite{Hawking:1995ap}.}

 {In contrast, our conjecture is that the canonical partition function $Z$ for the black hole horizon is given solely in terms of the boundary integral (\ref{RNboundary}). Let us check this: since} the energy at the horizon is
 $E(r=r_+) = m_S - 4\pi m_P^2 q^2/r_+ = m_S + U_e q_e/2 + U_b q_b/2$, the free energy, $F = E -\beta^{-1}S = m_S/2$, {is indeed directly obtained from the boundary term (\ref{RNboundary}) without any adjustments}.

%We impose that $U= \mp q_b/r_+$. We see it is the only choice, together with the $\alpha=\pm 1$, which leads to the consistent result (\ref{RN}).
%This calculation somewhat parallels the derivation of the canonical Noether charge $C$. We have
%\be
%C = -\frac{1}{4\pi }\int\diff^3 x\sqrt{-g}\lp F^{0\mu}+\alpha \tilde{F}^{0\mu}\rp n_\mu =  
%\frac{1}{4\pi }\int \diff r\diff^2\Omega r^2\lp \frac{q_e}{r^2} - \alpha\frac{q_b}{r^2}\rp = q_e - \alpha q_b\,. 
%\ee
%So, $q_b$ contributes to the usual charge formula only in the presence of the axion. In particular, each contribute to the charge on equal footing if $\alpha = \pm 1$. 

\section{Black hole in anti-de Sitter space}
\label{sec7sads}

For $f(r) = r_S/r +\Lambda r^2/3$, the horizon radius $r_h$ is given by the solution to the equation $f(r_h)=1$. For $\Lambda<0$, the solution is unique. We assume this, since the opposite case would be complicated by the presence of two horizons which are not in thermal equilibrium.  Now the surface temperature, $\beta^{-1} = -f'(r_h)/4\pi = (1-\Lambda r_h^2)/4\pi r_h$. 
%We integrate over the interior of the black hole, which amounts to change of signature of $(t_E,r)$ from $(+,+)$ to $(-,-)$ that does not formally change the derivation.
 The horizon boundary term gives
%\bs
\be \label{hboundary}
-\frac{m_P^2}{2}\int\diff^3 x\sqrt{g} V^\mu n_\mu = \frac{m_P^2}{2}\frac{4\pi r_h}{1-\Lambda r_h^2}\lp \frac{4}{3}\Lambda r_h + \frac{r_S}{r_h^2}\rp 4 \pi r_h^2  = 8\pi^2 m_P^2 r_h^2\frac{1+\Lambda r_h^2}{1-\Lambda r_h^2} = 2\pi m_P^2 \beta\lp r_h + \Lambda r_h^3\rp\,, 
\ee
where we solved $r_S$ from the equation $f(r_h)=1$. Now, since the energy is
\bs
\be
E(r=r_h)= m_S + \frac{4}{3}\pi m_P^2\Lambda r_h^3 = 4\pi m_P^2 r_h\,, 
\ee
and the temperature times entropy is
\be
\beta^{-1}S = \beta^{-1}(8\pi m_P^2) \pi r_h^2 = -2\pi m_P^2 \lp r_h - \Lambda r_h^3\rp\,, 
\ee
\es
we easily verify that now $I_E = \beta F$, where $F=E-\beta^{-1}S$ is the Helmholtz free energy of the black hole.  
%and the bulk term gives
%\be
%-\frac{m_P^2}{2}\int \diff^4 x_E\sqrt{g}\lp \mathring{R}-2\Lambda\rp = -\frac{16\pi^2 m_P^2 r_h^4\Lambda}{3\lp 1-\Lambda r_h^2\rp}\,,
%\ee
%\es
%and the sum yields us 

%This is the known result \cite{Hawking:1982dh}, even though 
The standard derivation relies on a completely different prescription \cite{Hawking:1982dh}: the boundary at the event horizon is not taken into account, but the integration is performed over the bulk $r_h < r < \infty$ and an asymptotic 3-dimensional boundary term plus a counterterm to the boundary are added at $r$ and then let $r \rightarrow \infty$. The counterterm has to be chosen
carefully in order to arrive at the established result,
\be \label{gibbs}
I^{(\text{GR})}_E = 2\pi m_P^2 \beta r_h\lp 1 + \frac{\Lambda r_h^2}{3}\rp\,.
\ee
As clarified in e.g. \cite{Dolan:2010ha}, this result corresponds to the partition function $Z$ of the Gibbs free energy $G$.  

The Gibbs free energy $G=\beta^{-1}I^{(\text{GR})}_E$ is finite since the enthalphy due to $\Lambda$ vanishes.  However, we expect the Helmholtz free energy to diverge. The total anti-de Sitter space is infinite and contains an infinite amount of (negative) energy\footnote{Finite results could perhaps be extracted by interpreting the $\Lambda$ as a part of gravity instead of a source term. The field equation $m_P^2\mathring{G}^\mu{}_\nu = \Lambda\delta^\mu_\nu$ reads $\mathring{\nabla}_\alpha H_a^{\alpha\mu} = {m_P^2}\Lambda\ie_a^\mu$ in the canonical form, but in an alternative $\Lambda$G$_\parallel$R theory it would read $\mathring{\nabla}_\alpha \tilde{H}_a^{\alpha\mu}=0$. In the latter theory, which could be interesting to investigate in the future, the zero-energy vacuum is an anti-de Sitter geometry. {This would align with the traditional intepretation of (\ref{gibbs}) as $I^{(\text{GR})}_E=\beta^{-1}\tilde{F}$, with the energy of $\Lambda$ excluded from $\tilde{F}=F-E^\Lambda$.}}, making our integral expressions divergent when considering the infinite space as a thermodynamical system. Let us rather consider the system up to some radius $r_0>r_h$ in the approach 1). Then, instead of (\ref{hboundary}) we consider the boundary at $r_0$, and instead of (\ref{GRbulk}) we integrate the bulk $r_h<r<r_0$. The sum of the two terms gives 
\be
I_E = 2\pi m_P^2\beta\lp r_S+\frac{4}{3}\Lambda r_0^3\rp - \frac{4}{3}\pi m_P^2 \beta\Lambda\lp r_0^3 - r_h^3\rp =   I^{(\text{GR})}_E + \beta E^\Lambda = \beta F\,,
\ee
which corrects the GR result by adding the energy $E^\Lambda = 4\pi m_P^2\Lambda r_0^3/3$ associated with $\Lambda$ contained within the radius $r_0$. It is satisfactory indeed that the $Z$ as defined in (\ref{partition}) is now interpreted as the canonical partition function for the Schwarzschild-anti-de-Sitter spacetime.

\section{Conclusions}
\label{sec8conc}

We derived partition functions for static spherically symmetric black holes applying the Euclidean path integral approach to G$_\parallel$R theory. 1) Following conventional approach, we computed the partition function for the total spacetime, but considered also the 2) quasilocal description of the black hole horizon. In the former case, for some practical purposes our proposal boils down to replacing the York-Gibbons-Hawking term for $\mathring{R}$ by a different boundary term in the action. Even then, ambiguities of the conventional derivations are avoided and, in particular, the new boundary term does not require the conventional counterterms. Further, the action always corresponded to the canonical black hole partition function, unlike in GR.
The latter approach 2) suggested a possible further improvement. The difference became apparent in the case of a charged black hole and in the presence of a cosmological constant. The quasilocal prescription led to the expected expressions for the Helmholtz free energy $F$, again without ad hoc adjustments of the action or re-interpretations of the results. 

This motivates pursuing the Euclidean approach in G$_\parallel$R theory further, beyond the on-shell approximation. It remains to be seen whether the computations are facilitated in this theory and how the results differ from those in GR at higher orders. It can also be worthwhile to check the validity of the prescriptions 1) and 2) in more general spacetimes than static spherically symmetric black holes.

To conclude, our results offer strong grounds to conjecture that the more robust and consistent formulation of Euclidean quantum gravity could be realised in the framework of G$_\parallel$R. %even nonperturbatively. 
The fact that the bulk action vanishes by construction in an inertial frame suggests that the Euclidean action could now be bounded from below. Moreover, the absence of counterterms in our computations indicates that quantum G$_\parallel$R could be renormalisable, or at least substantially better behaved in the ultraviolet than standard GR.

%The fact that the computations required no counterterms in an inertial frame suggests that the quantum G$_\parallel$R may be renormalisable, or perhaps even finite.

%G$_\parallel$R is a radically new (or perhaps rather, a radically old  \cite{Koivisto:2022nar}) theory of relativity.    
%The gauge part $A^\pm_\mu \rightarrow A^\pm_\mu -  (t q_e/r_\pm^2)_{,\mu}$ is required because $t$ is ill-defined at the horizon \cite{Gibbons:1976ue}. The on-shell action (\ref{maxwell2}) for the solution (\ref{EMsolution}) is $I_M=0$. 

\acknowledgements{TSK was supported by the Estonian Research Council grants CoE TK202 “Foundations of the Universe” and PRG2608 “Space - Time  - Matter”. J.B.J. was supported by the Project PID2021-122938NB-I00 funded by the Spanish “Ministerio de Ciencia e Innovaci\'on" and FEDER “A way of making Europe” and the Project SA097P24 funded by Junta de Castilla y Le\'on.}

\bibliography{paraeuclid}

\end{document}